\documentclass[aip,apl,
amsmath,amssymb,
 reprint,
]{revtex4-2}

\usepackage{graphicx}
\usepackage{dcolumn}
\usepackage{bm}

 \usepackage{xspace}

\usepackage[utf8]{inputenc}
\usepackage[T1]{fontenc}

\graphicspath{{active_figs/}}

\newcommand{\vecr}{\boldsymbol{r}}

\newcommand{\vecu}{\boldsymbol{u}}
\newcommand{\vecU}{\boldsymbol{U}}

\newcommand{\micron}{~\textmu m\xspace}

\newcommand{\modeprofile}{\tilde{\vecu}(x, y)}
\newcommand{\dx}{\mathrm{d}x}
\newcommand{\dy}{\mathrm{d}y}
\newcommand{\neff}{n_{\text{eff}}}

\begin{document}

\preprint{APL Photonics}

\title{Anti-resonant reflecting acoustic rib waveguides for strong opto-acoustic interaction}

\author{Thomas J. Dinter}
\author{Miko\l{}aj K. Schmidt}
\author{Michael J. Steel}

\affiliation{School of Mathematical and Physical Sciences, Macquarie University, NSW 2109, Australia}
\email{michael.steel@mq.edu.au}

\date{\today}

\begin{abstract}
Few known material systems can simultaneously guide optical and elastic fields through total internal reflection. This natural limit has restricted the realization of strong optoacoustic effects to highly-specialised and purpose-built platforms which employ either  exotic materials, or complex waveguide designs. Here we apply the concept of Anti-Resonant Reflecting Acoustic Waveguides (ARRAWs) as a potential solution to this issue. ARRAWs confine the elastic field to a high-elastic-velocity core via the anti-resonances of a cladding layer of lower elastic velocity. We numerically study the appearance and dispersion of ARRAW-guided modes in a conventional silicon-on-insulator rib waveguide geometry. Applying the technique to the problem of efficient backwards Stimulated Brillouin Scattering (SBS), we predict that ARRAW guidance, in conjunction with conventional optical confinement, can produce Brillouin gains comparable to those of more exotic geometries.
\end{abstract}

\maketitle

\section{Introduction}
Achieving efficient co-localization of  optical and elastic fields in photonic waveguides is important for enhancing  on-chip nonlinear optoacoustic interactions, notably  Stimulated Brillouin Scattering (SBS) and its wide range of potential applications in integrated photonic circuits~\cite{pant_chapter_2022, pant_-chip_2011}, particularly in the domain of microwave photonics~\cite{CKLai2024, marpaung_integrated_2019}. Thus one needs to identify a combination of physical effects that, as far as possible, can trap light and sound waves in the same spatial region.

\begin{figure}[ht!]
    \begin{center}
        \includegraphics[width=\linewidth]{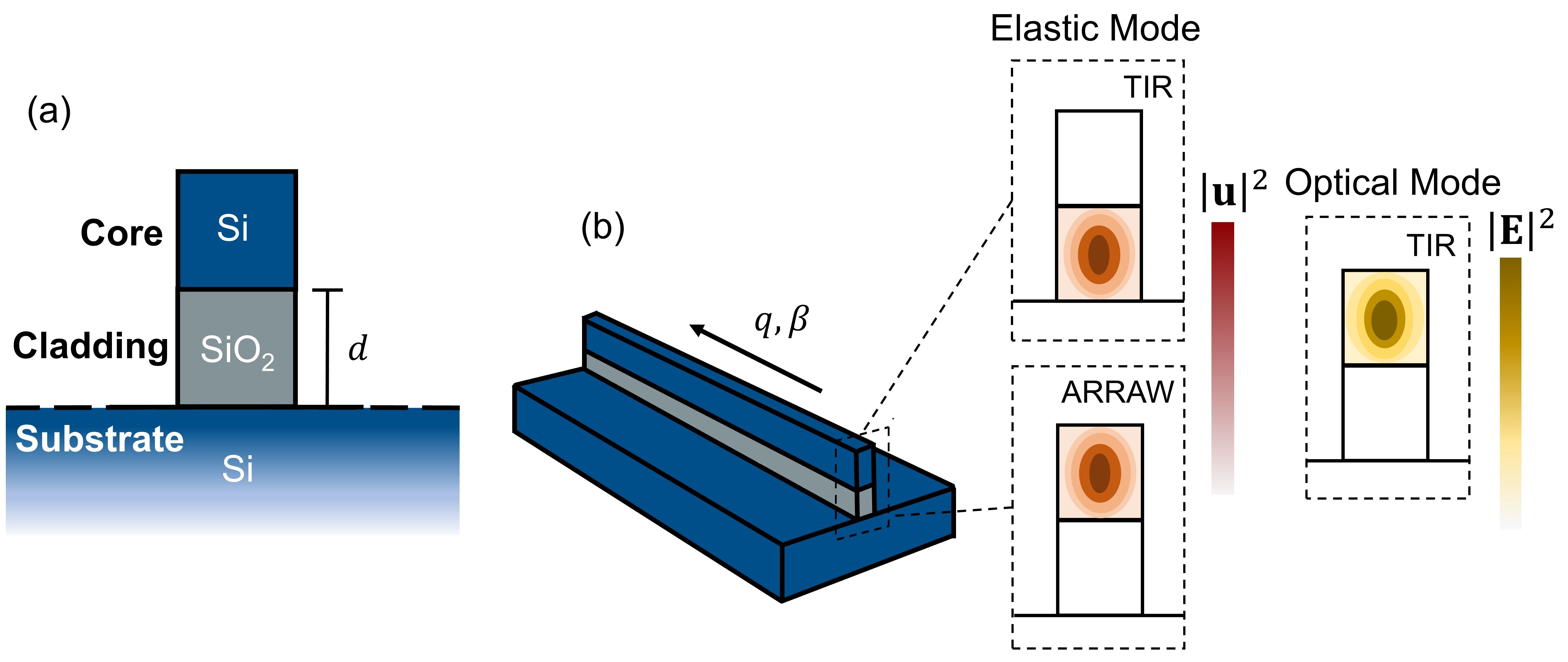}
        \caption{
            (a) Schematic of a longitudinally-invariant Si/SiO$_{2}$ rib waveguide. (b) Optical modes are conventionally guided in the high-index silicon core via TIR (far right inset). Conversely, elastic modes can be either conventionally guided in the silica cladding (upper inset) resulting in low spatial overlap between the two fields, or anti-resonantly guided in the core (lower inset) with the potential for strong optoelastic coupling.
        }
        \label{fig:WaveguideDrawing}
    \end{center}
\end{figure}

The most obvious strategy to achieve simultaneous confinement of light and sound is to seek waveguide platforms in which both the optical and elastic fields can be guided in the waveguide core through total internal reflection (TIR).
However few material systems are compatible with this approach, since most optical materials with a high refractive index also exhibit high material stiffness~\cite{eggleton_brillouin_2019}. Elastic phase velocities have the general form $v=\sqrt{c~/~\rho}$ for some stiffness coefficient $c$, where the density $\rho$ exhibits relatively little variation between dielectric materials.  The high index core of a conventional optical waveguide thus typically exhibits elastic velocities \emph{larger} than in the nominal cladding layers, and so joint light/sound TIR is not possible. Consequently, there is minimal spatial overlap between optical and elastic fields in standard dielectric waveguide designs, and therefore negligible optoelastic nonlinearity (see Fig.~\ref{fig:WaveguideDrawing}).

Several approaches have been explored to address this problem. Perhaps the most common strategy is to partially isolate the acoustic waveguide from the substrate, by under-etching the system. The thin supporting structures~\cite{shin_tailorable_2013, kittlaus_large_2016, laer_net_2015, schmidt_suspended_2019} still couple the elastic field to the substrate but, with careful engineering, this radiative leakage can be made  small compared to the intrinsic viscous losses. Confinement in periodic structures with line defects, as is the case in phoxonic crystals~\cite{maldovan_simultaneous_2006, zhang_design_2017, yu_giant_2018}, is another common approach with a similar mechanism.

Yet another strategy is to adopt one of the small number of  established waveguide platforms that \emph{do} support co-localized TIR guidance for both fields. The best-known example is the chalcogenide chip family~\cite{pant_-chip_2011, eggleton_chalcogenide_2011, lai_hybrid_2022}, with more recent candidates including lithium niobate (LNOI)~\cite{wang_tailorable_2021, feng_intramode_2022, yu_gigahertz_2021} and silicon nitride~\cite{wang_design_2020, botter_guided-acoustic_2022, botter_ra_stimulated_2021, gyger_observation_2020}. However, while these present promising platform-specific solutions, it is also desirable to consider material-agnostic approaches which would allow  the use of standard dielectric optical platforms, facilitating fabrication and integration into existing systems.

Recently, we introduced such an alternative approach in the form of Anti-Resonant Reflecting Acoustic Waveguides (ARRAWs)~\cite{schmidt_arraw_2020}, whereby  GHz elastic modes are confined to a high-elastic-velocity core using the Fabry-Perot anti-resonances of a low-elastic-velocity cladding layer~\cite{schmidt_arraw_2020}.
This concept was developed in analogy to  Anti-Resonant Reflecting Optical Waveguides (ARROWs), initially conceived in low-contrast planar optical circuits in the 1980s~\cite{archambault_loss_1993, duguay_antiresonant_1986}, and then further developed in the context of microstructured optical fibers in the early 2000s~\cite{Litchinitser:02,Litchinitser:03,White:02,Steinvurzel:04}.
In our work, we theoretically studied a selection of idealised cylindrical and planar geometries, finding regimes of efficient guidance, which consequently showed strong Brillouin gain.  We also found that due to the competing contributions of both shear (S) and longitudinal (P) polarizations, which have different phase velocities, the elastic ARRAW problem is inherently more subtle and challenging to engineer than the original ARROW concept which addresses optical confinement for which there are of course no longitudinal modes.

Very recently, Lei~\emph{et al}~\cite{lei_anti-resonant_2024} have extended this idea to a suspended under-etched waveguide design, where the elastic field is confined in the horizontal direction by anti-resonances associated with a sequence of deeply-etched side trenches. In their experiments, for forward Brillouin scattering, they show impressive Brillouin gains, in one case exceeding 3500~$(\mbox{Wm} )^{-1}$. For backward SBS, they report gains of order 400-600~$(\mbox{Wm} )^{-1}$, still significant. It is then interesting to ask what degree of confinement and Brillouin gains might be obtained using ARRAW confinement in realistic structures but without the fabrication and mechanical complexities of suspended under-etched designs like in Ref.~\cite{lei_anti-resonant_2024}.

In this work, we study the potential for ARRAW guidance in a family of simple rib-like  waveguides,  with the anti-resonance condition designed to achieve \emph{vertical} confinement, rather than the more familiar horizontal confinement. An example  configuration for the silicon-on-insulator platform is shown in Fig.~\ref{fig:WaveguideDrawing}. We determine the elastic eigenmodes of a structure in the GHz domain, and use a combination of filtering criteria to construct the dispersion relation of the anti-resonant modes. Further, we quantify the extent of confinement possible through the optimization of the mechanical quality factor $Q_{\text{m}}$, and predict useful backward SBS gains for a range of interacting modes.

\section{ARRAW guidance}\label{IdealisedARRAWCondition}

In our previous work introducing  anti-resonant acoustic guidance,
we concentrated on ideal cylindrical and planar geometries, and found families of elastic modes in which the elastic displacement field is confined to a high-elastic-velocity core via anti-resonances in the cladding layer~\cite{schmidt_arraw_2020}.
Formally, the anti-resonant modes are leaky resonances with a non-zero longitudinal decay rate. However with appropriate designs, we found the leakage rate can be reduced to well below the intrinsic material viscous losses of the structure that apply in the GHz frequency regime.

By analogy to the condition for optical anti-resonances~\cite{Litchinitser:02}, one might expect these resonances should satisfy the relation
\begin{align}\label{3D_ARRAW_Condition}
    k_{\perp}^\sigma d & = m^\sigma\pi + \frac{\pi}{2}\,,
\end{align}
for cladding width $d$, transverse wavenumber  $k_{\perp}^\sigma$ and modes indexed by integers $m^\sigma$, where $\sigma$ denotes P (longitudinal) or S (shear) polarization. In the optical case with simple cylindrical or planar geometries, the analogous TM and TE polarization states are separable and Eq.~\eqref{3D_ARRAW_Condition} holds straight-forwardly. In the elastic case however, the waveguide interfaces couple the S- and P-waves in a non-trivial fashion. Consequently, the two anti-resonance conditions cannot be satisfied simultaneously, limiting the applicability of Eq.~\eqref{3D_ARRAW_Condition} for even relatively simple waveguide geometries. In our initial work~\cite{schmidt_arraw_2020}, we addressed this problem by considering modes with only one polarization component (such as the pure torsional modes in cylindrical waveguides), or by relaxing the above conditions. In the latter case, we characterized the efficacy of the ARRAW mechanism in an operational fashion, by searching for the local minima of the leakage rate as a function of wavelength or cladding width (or equivalently, by searching for the maxima of a leaky resonance's mechanical quality factor $Q_m$, defined below). We  adopt a similar approach in this work applied to more practical structures.

\subsection{Modal analysis}
To identify the dispersion of the anti-resonant modes in more complex waveguide geometries, such as the rib waveguide shown in Fig.~\ref{fig:WaveguideDrawing}, we search for numerical solutions of the full elastic wave equation.
Written in component notation, the elastic wave equation for the real-valued time-dependent displacement field $\vecU(\vecr, t)$
is~\cite{auld_acoustic_1973}
\begin{align}\label{WaveEquation}
    \frac{\partial}{\partial r_{j}}\left(c_{ijkl}S_{kl}+\eta_{ijkl}\frac{\partial S_{kl}}{\partial t}\right)
     & = \rho(\vecr)\frac{\partial^{2}U_{i}}{\partial t^{2}}\,.
\end{align}
Here $\rho(\vecr)$ is the mass density,
\begin{align}
    S_{ij}(\vecr, t)=\frac{1}{2}\left(\frac{\partial U_i}{\partial r_j}+\frac{\partial U_j}{\partial r_i}\right) ,
\end{align}
is the symmetric strain tensor,  $\overleftrightarrow{c}$ and $\overleftrightarrow{\eta}$ are the fourth rank stiffness and viscosity tensors, respectively. Repeated subscripts in Eq.~\eqref{WaveEquation} are summed over the coordinates $x, y, z$. In this work, we restrict the discussion to piecewise-constant structures composed of materials which we label with a superscript $(n)$. All values of the material properties are provided in Appendix~\ref{MaterialConstants}.

Seeking solutions with harmonic time dependence
\begin{equation}
    \vecU(\vecr,t) = \vecu(\vecr)\exp(-i\Omega t)+\vecu(\vecr)^*\exp(i\Omega t),
\end{equation}
and adopting lossless media such that $\overleftrightarrow{\eta}^{(n)}=0$, we may simplify Eq.~\eqref{WaveEquation} to
\begin{align}\label{EigenEquation}
    -\frac{\partial}{\partial r_{j}}\left(c_{ijkl}^{(n)}\frac{\partial u_{k}}{\partial r_{l}}\right) - \rho^{(n)}(\vecr)\Omega^{2}u_{i} = 0\,.
\end{align}
More specifically, for a waveguide longitudinally-invariant along the $z$-axis, we seek modal solutions of the form $\vecu(\vecr)=\tilde{\vecu}(x,y)\exp(i q z)$ where $q$ is the elastic propagation constant. To account for the leaky nature of the resonances, at least one of $\Omega$ and $q$ must be allowed to be complex. In this work, we treat the wavenumber $q$ as the real independent variable, and the angular frequency $\Omega(q)$ as the complex eigenvalue, with the dissipation implying $\text{Im}(\Omega)<0$.

\begin{figure}[ht!]
    \begin{center}
        \includegraphics[width=\linewidth]{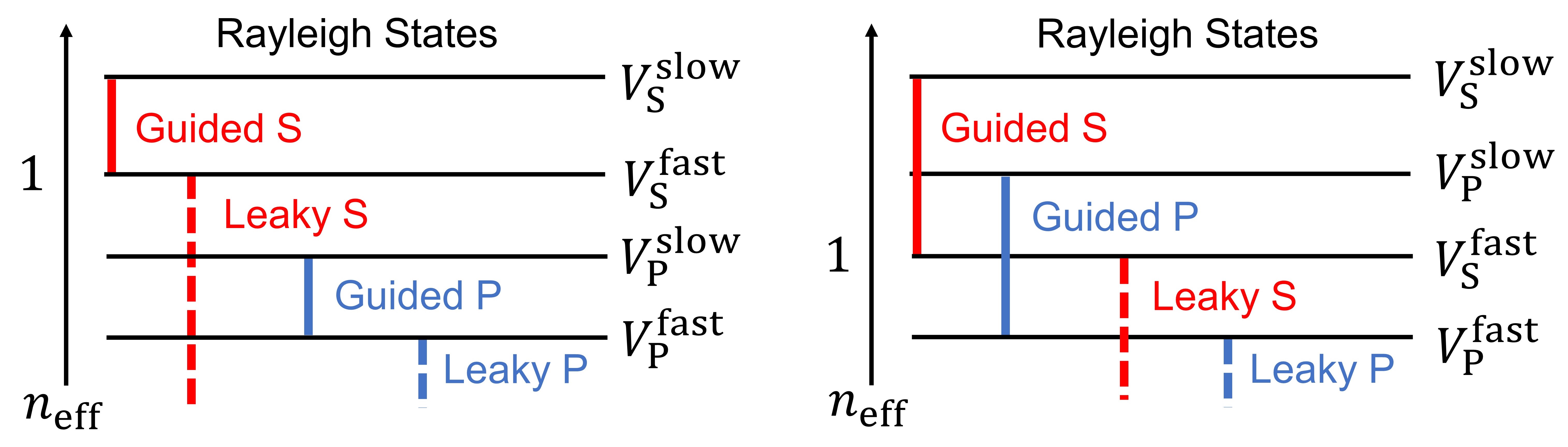}
        \caption{
            The two velocity orderings possible for an elastic waveguide, with the behaviour of shear- and longitudinally-polarised modes in each region of the effective index plane labeled accordingly. Examples of material systems possessing these velocity orderings include Si/SiO$_{2}$ (left), and Si/As$_{2}$S$_{3}$ chalcogenide (right) platforms. 
        }
        \label{fig:VelocityOrdering}
    \end{center}
\end{figure}

Considering isotropic materials, each material has a pair of intrinsic P and S-wave phase velocities, with $v_S< v_P$.
Depending on the choice of core  and cladding   materials used for the waveguide model in Fig.~\ref{fig:WaveguideDrawing}, there are thus two possible orderings of the elastic velocities, as illustrated in Fig.~\ref{fig:VelocityOrdering}. Since there is considerable risk of confusion, we remark once again that the labels ``core'' and ``cladding'' are assigned in terms of the optical refractive index, and so the elastic properties are the reverse of what we would expect in a TIR configuration: here the ``core'' is elastically fast and the ``cladding'' is elastically slow, as denoted in Fig.~\ref{fig:WaveguideDrawing}(a).

The two columns in Fig.~\ref{fig:VelocityOrdering} differ in the ordering of the two inner speeds: the S-wave of the ``fast'' core layer and the the P-wave of the ``slow'' cladding layer.  Here, we concentrate on the  ordering characteristic of Si/SiO$_2$ platforms, for which the S-wave velocity in Si ($V_{\text{S}}^{\text{fast}}$) exceeds that of the P wave in SiO$_2$ ($V_{\text{P}}^{\text{slow}}$). That is, $V_{\text{S}}^{\text{slow}}~\leq~V_{\text{S}}^{\text{fast}}~\leq~V_{\text{P}}^{\text{slow}}~\leq~V_{\text{P}}^{\text{fast}}$.

We also define an effective mode index $n_{\text{eff}}$ in terms of the modal phase velocity $V_{p}~\equiv~\text{Re}(\Omega)~/~q$ as
\begin{align}\label{neff}
    n_{\text{eff}} = \frac{ V^{\text{fast}}_{\text{S}}}{V_{p}}\,.
\end{align}
The choice of $V^{\text{fast}}_{\text{S}}$ as the reference velocity in this definition is arbitrary, but results in a convenient lower bound for the effective index of conventionally (TIR) guided elastic modes of $n_\text{eff}=1$. Proceeding vertically downwards in Fig.~\ref{fig:VelocityOrdering},
Rayleigh-like surface states may be expected in the range $\neff>n_R$, where ${n_R = V^{\text{fast}}_{\text{S}}~/~V^{\text{slow}}_{\text{S}}}$.
Next, all conventionally guided modes are found in the region $1\leq n_{\text{eff}} \leq n_R $, and leaky modes for $n_{\text{eff}}<1$. Note that due to our choice of velocity orderings, as in our previous work~\cite{schmidt_arraw_2020}, only modes with pure shear character can be conventionally guided.

Most modes of interest will thus be leaky, and best characterised as spectral resonances with a linewidth $\text{Im}(\Omega)$. We quantify the leakage loss of any given mode by assuming a real propagation constant $q$, and using the complex frequency to define a mechanical quality factor ${Q_{\text{m}}=\text{Re}(\Omega)~/~(2\text{Im}(\Omega))}$. To account for the effects of material losses on these quality factors, we have assumed an additive model of losses such that ${1/Q=1/Q_{\text{m}}+1/Q_{\text{v}}}$, where $Q_{\text{m}}$ refers to the purely radiative quality factor discussed thus far, and $Q_{\text{v}}$ accounts for dissipation due to the intrinsic viscosity. Setting $Q_{\text{v}}$ to a fixed value of $10^{3}$ therefore places a realistic upper-limit on the performance of our system.

\begin{figure}[ht]
    \begin{center}
        \includegraphics[width=\linewidth]{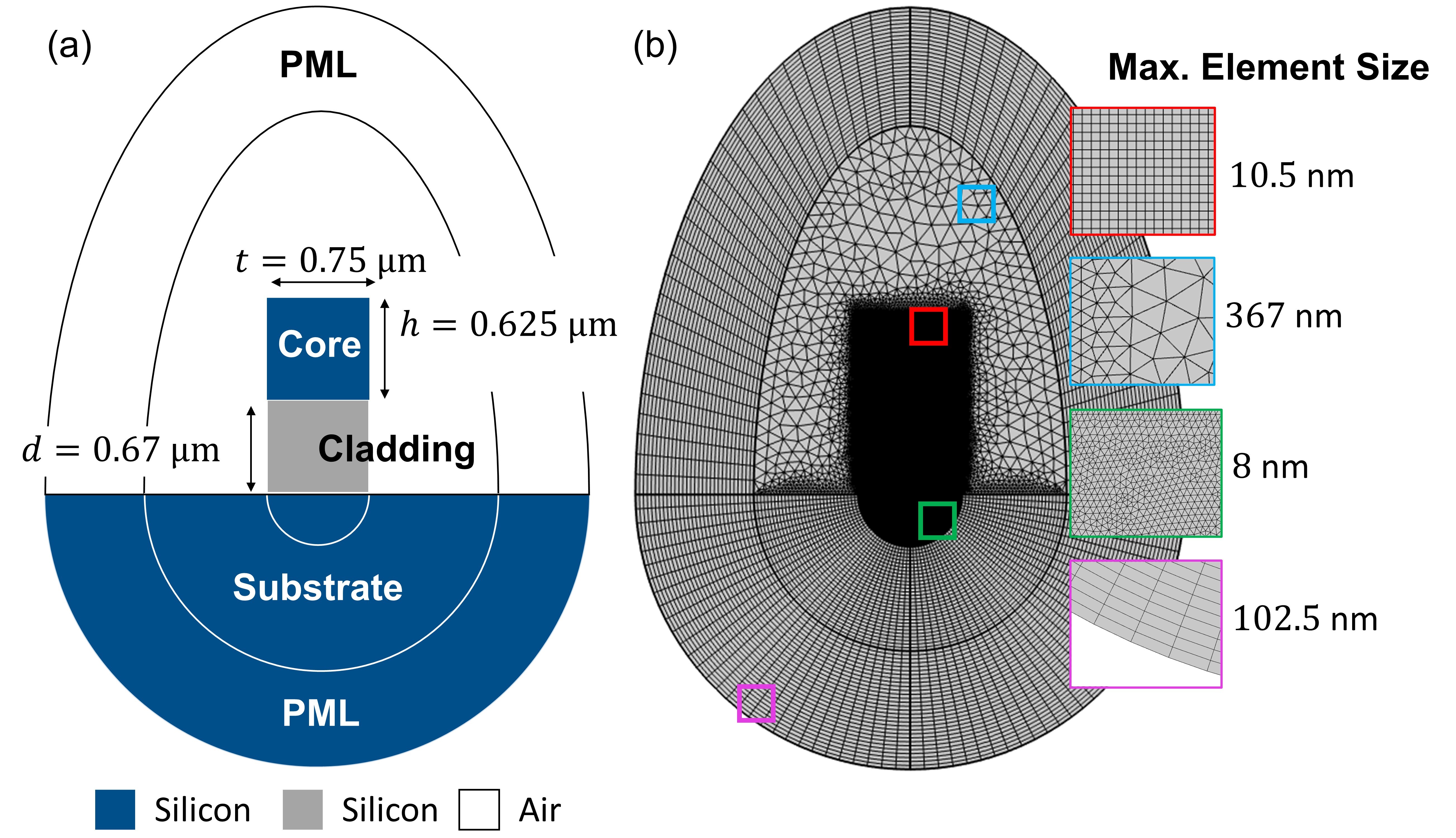}
        \caption{
            \textbf{(a)} The full calculation domain used for all finite element simulations, with key geometric parameter values shown. Here, the outermost layers are composed of air and silicon (Si) respectively, and are set as perfectly matched layers (PMLs). \textbf{(b)} The mesh used in COMSOL for finite element simulation. A short discussion on meshing details can be found in Appendix \ref{MeshingDetails}.
        }
        \label{fig:SimDomain}
    \end{center}
\end{figure}

\begin{figure*}[ht] 
    \begin{center}
    \includegraphics[width=1\linewidth]{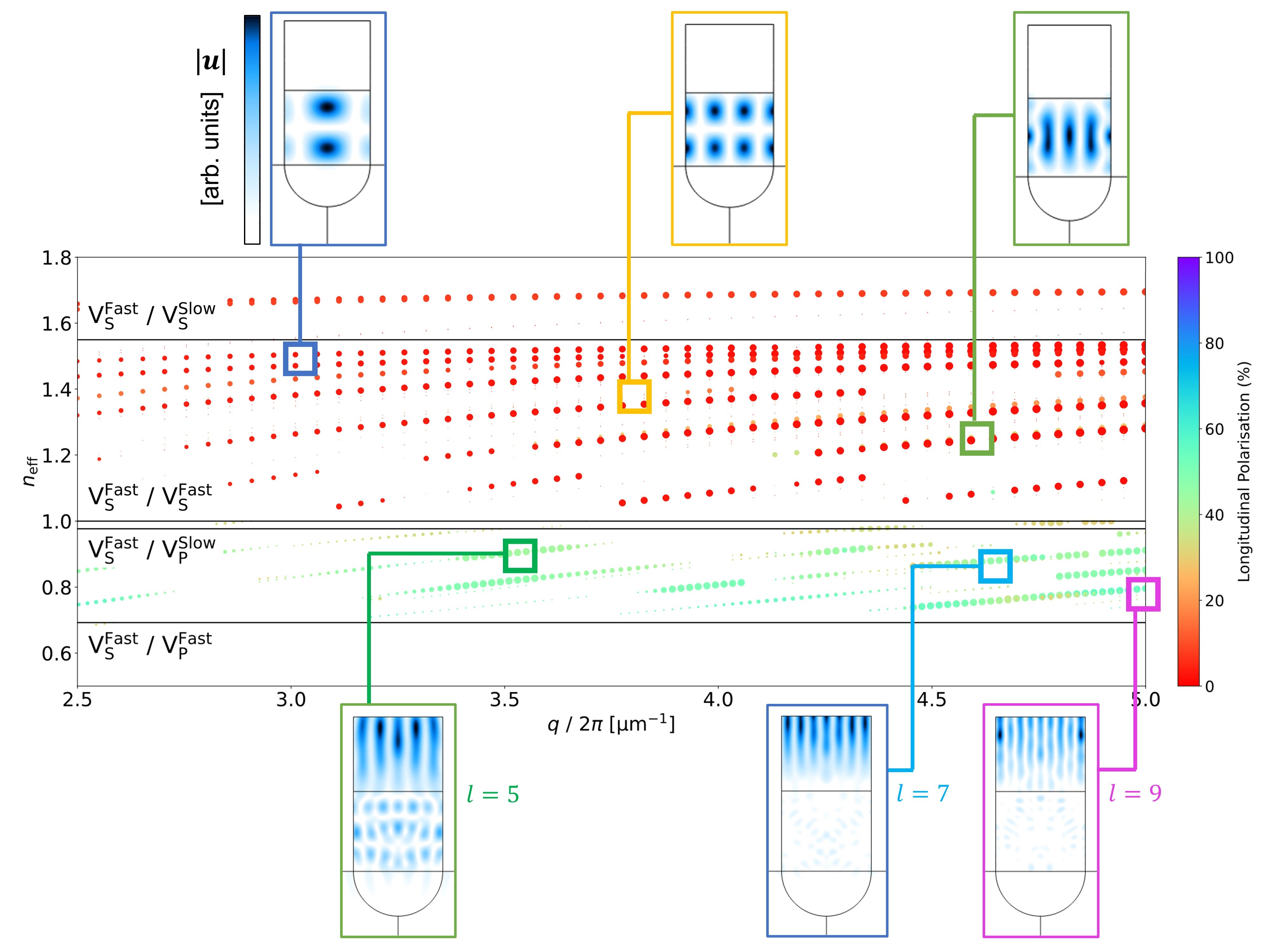}
    \caption{ 
    Numerical calculation of 
    elastic dispersion diagram of a longitudinally-invariant
    Si/SiO$_{2}$ rib waveguide with dimensions provided in
    Fig.~\ref{fig:SimDomain} above. For the conventionally guided modes with $n_{\textbf{eff}}> 1$,
     the size of each point is inversely
    proportional to the second moment width of the mode defined in
    Eq.~(\ref{SecondMomentWidth}), Select mode profiles are attached as
    insets above the main plot. For the region with $n_{\textbf{eff}}< 1$,
    the size of each point is dictated by the localization value of the 
    mode, defined in Eq.~(\ref{UnconventionalFilter}).  
    Mode profiles for the ARRAW modes with $l=5,~7$ and $9$ modes are shown as insets.
} 
    \label{fig:Dispersion} \end{center}
    \end{figure*}

\section{Classifying elastic resonances of the rib}

We now explore in detail the elastic dispersion properties of a rib waveguide of the general form of Fig.~\ref{fig:WaveguideDrawing} with the particular parameters shown in Fig.~\ref{fig:SimDomain}.
The calculations were performed using the finite element method (FEM) simulation tool COMSOL (see Appendices~\ref{NumericalSolverDetails} and~\ref{MeshingDetails} for details on the mesh and other numerical settings).
We assume a silicon (Si) core height of $h = 0.625$\micron and a silica (SiO$_2$) cladding height of $d= 0.67$\micron, both of which are $t= 0.75$\micron in width. The structure rests on a Si substrate and the physical simulation domain is surrounded by a FEM perfectly matched layer (PML)~\cite{berenger_perfectly_1994, qi_evaluation_1998}. Note that these dimensions are chosen arbitrarily, and are comparable to typical SoI platforms.  The possibly surpr  ovoid shape of the simulation domain is designed to assist the efficacy of the FEM PML.


The resulting elastic dispersion map is shown in Fig.~\ref{fig:Dispersion}. We can identify several families of eigenstates: Rayleigh-like surface states with $n_{\text{eff}}> V^{\mathrm{fast}}_{\mathrm{S}}~/~V^{\mathrm{slow}}_{\mathrm{S}}$, conventionally guided modes with ${1\leq n_{\text{eff}}\leq V^{\mathrm{fast}}_{\mathrm{S}}~/~V^{\mathrm{slow}}_{\mathrm{S}}}$, and {leaky} resonances with $n_{\text{eff}}<1$. The color of each point in the diagram encodes the relative contribution of the longitudinal polarization to the elastic mode, defined as $\int |\tilde{u}_{z}(x, y)|^{2}\,\mathrm{d}A~/~\int|\modeprofile|^{2}\,\mathrm{d}A$.
The size of each point indicates the degree of localization of the field to the cladding (for the conventionally guided and Rayleigh modes), or the core (for the leaky modes).  Larger points indicate stronger confinement (see Eq.~\eqref{UnconventionalFilter}). The apparent discontinuities in some dispersion curves are due to additional filtering conditions described below, implemented to discard resonances  with low quality $Q_{\text{m}}$, and spurious non-physical modes identified by the numerical solver. We focus on the wavenumber range
${2.5\mbox{\micron}^{-1} \leq q~/~(2\pi) \leq 5\mbox{\micron}^{-1}}$ in which we find good confinement of the $\neff< 1 $ mode families.


The upper band in the range $\neff> 1.55$ exhibits a family of Rayleigh-like states confined close to the surface of the silicon layer, with characteristically weak dispersion~\cite{auld_acoustic_1973}. For this structure, the Rayleigh states do not bring the elastic and optical fields into close contact and they will not be discussed further. We note that some recent studies in the thin-film lithium niobate platform has shown these kinds of states can have applications in SBS~\cite{Rodrigues:23,rodrigues2023b,ye2023}.

Next, we find a set of conventionally guided elastic modes ($1 \leq n_{\text{eff}} \lesssim 1.55$), characterised by predominantly shear polarization (red-orange dots), and mode profiles which are localized to the low-elastic-velocity cladding region via total internal reflection. Three examples with different patterns of vertical and horizontal nodes are shown in the upper panel of Fig.~\ref{fig:Dispersion}. Appendix~\ref{app:tirmodes} discusses some technical points on the procedure for identifying these states.



\begin{figure}[ht!]
    \begin{center}
        \includegraphics[width=\linewidth]{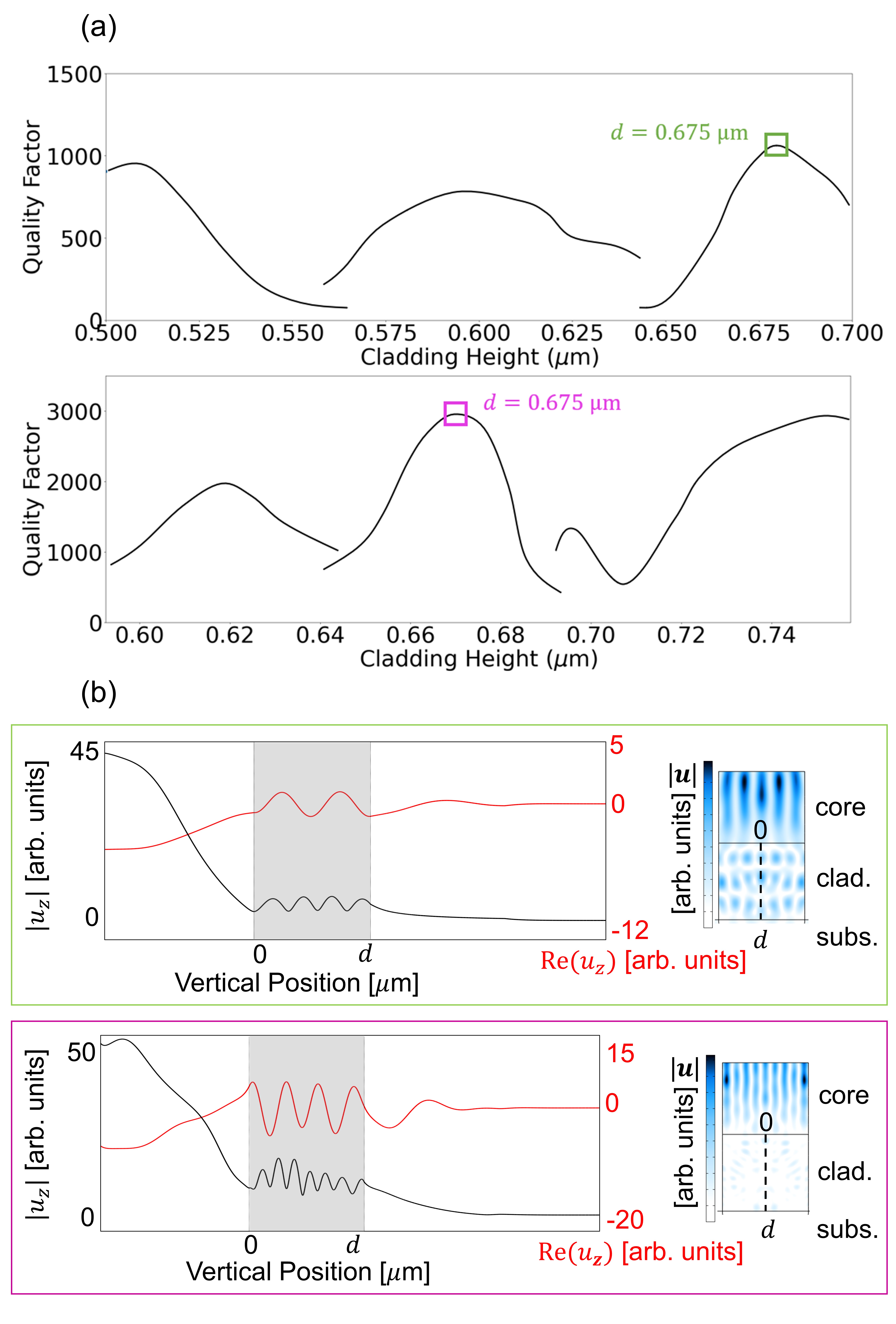}
        \caption{
            \textbf{(a)} The mechanical quality factor $Q_{\mathrm{m}}$ of the 5th (upper) and 7th (lower) order anti-resonant elastic modes, for varying cladding height~$d$. \textbf{(b)} The displacement profiles of the corresponding modes labeled in~(a) as a function of vertical position (left). The corresponding mode profile $\tilde{\boldsymbol{u}}(x, y)$ is also shown (right).
        }
        \label{fig:EvidenceForARRAW}
    \end{center}
\end{figure}


Finally, we discuss the modes in question: leaky resonances in the spectral region $n_{\mathrm{eff}} < 1$. When performing the calculations, since we are seeking leaky modes, as well as resonances which are localised to the waveguide, the FEM solver also typically identifies a large number of spatially-extended quasi-continuum radiation states which are of little physical interest but which fill up the spectrum. To extract the dispersion relation of just the leaky states localised to the core, we filter all the states by a measure of their core confinement.
The selection criterion is expressed as
\begin{equation}\label{UnconventionalFilter}
    \frac{\int_{\text{Core}}|\modeprofile|^{2}\,\mathrm{d}A}{\int_{\text{Total}}|\modeprofile|^{2}\,\mathrm{d}A} \geq \gamma\,,
\end{equation}
for some $\gamma \in [0,1]$. We found a value of $\gamma=0.7$ was effective in eliminating most of the radiation states.

Once this filtering condition is applied, we are left with a distinctive set of bands colored green-yellow which are identified as the leaky modes strongly localized to the core. As illustrated in the bottom row of insets in Fig.~\ref{fig:Dispersion}, these modes are predominantly localised to the desired core region. Once again, the size of each point is correlated with its degree of confinement to the core.
We have labeled these modes by the number $l$ of horizontal oscillations of the field in the core.

There are several interesting features in the dispersion map.
In contrast to the transverse nature of the cladding-confined modes discussed above, these modes have a mixed character with significant longitudinal contribution (observe the green/yellow hue of these bands). This is significant, as the polarization of an elastic mode partially determines its suitability for use in Brillouin scattering processes, with BSBS in particular relying on a strong P-wave component of the acoustic modes~\cite{wolff_brillouin_2021}.


Secondly, as indicated by the size of the dots representing the mode confinement, the bands appear to fade in and out several times as the wavenumber $q$ changes.
This oscillatory character is strongly distinctive of an ARRAW mechanism of guidance associated with anti-resonances of the low-velocity cladding layer.
In our previous work~\cite{schmidt_arraw_2020}, we found that the radiative mechanical quality $Q_{\mathrm{m}}$ displayed just these kinds of oscillations as a function of the cladding width or elastic wavenumber.
In Fig.~\ref{fig:EvidenceForARRAW}(a) we show the dependence of the quality factor of the large-$Q_{\mathrm{m}}$ modes on the cladding height calculated along the dispersion curves of the  modes marked as $l=5$ and $l=7$ in Fig.~\ref{fig:Dispersion}, with the same character of a succession of resonances being apparent. Each maximum of $Q_{\mathrm{m}}$ corresponds to an ARRAW resonance attaining its minimum leakage and maximum confinement.

Another feature of the anti-resonant behaviour of these modes is shown in Fig.~\ref{fig:EvidenceForARRAW}(b), where we plot the displacement profiles $\Re\big({u}_{z}(y)\big)$ (red lines) and $|{u}_{z}(y)|$ (black) along the vertical profile of the waveguide, and show that the former exhibit approximately $3.5\pi$ and $6.5\pi$ phase evolution within the cladding. The considerable confinement of the field to the inner core region on the left of each plot is also clear.

We end this section with a brief remark on the amended criterion for ARRAWs used throughout this work, compared to Ref.~\cite{schmidt_arraw_2020}. In that original contribution, we proposed that ARRAW modes should be found in the region $n_{\mathrm{eff}} < v_\text{S}^\text{fast}~/~v_\text{P}^\text{fast}$ --- a criterion introduced as a direct parallel of the condition for optical ARROWs $n_{\mathrm{eff}} < c~/~v^\text{fast}$. Here however, we find that in the much broader region $n_{\mathrm{eff}} < 1$, we can identify modes with the distribution (see Fig.~\ref{fig:Dispersion}), and resonant character of dissipation (see Fig.~\ref{fig:EvidenceForARRAW}) indicating the ARRAW mechanism of guidance. Furthermore, all the leaky modes found in this region will have nonvanishing real components of the transverse wavenumbers, meaning that they can approximately realise the anti-resonance condition. We therefore argue that the condition for ARRAWs given in Ref.~[\citenum{schmidt_arraw_2020}] should be relaxed, to incorporate all the relevant modes in the $n_{\mathrm{eff}} < 1$ region. 


\section{Application to backwards SBS}

We now consider the use of these modes in the context of Stimulated Brillouin Scattering (SBS), specifically backward SBS (BSBS). In characterizing BSBS we seek the gain coefficient $\Gamma$ describing the growth of the backward-propagating Stokes (S) optical mode subject to a forward-propagating pump (p), the interaction being mediated by a spontaneously generated phase-matched elastic field. In the undepleted pump regime,  the power in the Stokes mode grows exponentially as $P_\text{S}(z)=P_\text{S}(L)e^{\Gamma P_\text{p,0}(L-z) }$, where $L$ is the waveguide length, $P_\text{p,0}$ is the input pump power and the gain coefficient is given by \cite{wolff_stimulated_2015}
\begin{align}\label{GainCoefficient}
    \Gamma = 4\omega_{\text{p}}\frac{Q_{\text{m}}}{\mathcal{P}_{\text{p}}\mathcal{P}_{\text{S}}\mathcal{E}_{\text{m}}}|\mathcal{Q}_{\text{SBS}}|^{2}\,.
\end{align}
Here $\omega_{\text{p}}$ is the angular frequency of the optical pump mode, and $\mathcal{Q}_{\text{SBS}}$ is the coupling strength of the optoacoustic interaction. The factors in the denominator account for modal normalization with  $\mathcal{P}_{i}$ being the energy flux of the optical pump and Stokes modal field, and $\mathcal{E}_{\text{m}}$ being the elastic energy density. (See Appendix~\ref{BSBSCalcs} for details).
Here we focus on \emph{intramodal} BSBS, where the pump and Stokes fields  occupy the same mode of the waveguide. The coupling term can be further decomposed as
\begin{align}\label{CouplingCoef}
    \mathcal{Q}_{\text{SBS}} = \mathcal{Q}^{\text{PE}} + \mathcal{Q}^{\text{MB}}\,,
\end{align}
where PE and MB refer to the photo-elastic and moving boundary effects, respectively. The need for the co-localization of optical and elastic modes enters through these terms, which take the form of weighted spatial-overlap integrals as discussed in Appendix~\ref{BSBSCalcs}. We have used Eq.~(\ref{GainCoefficient}) to determine gain coefficients $\Gamma$ for the first seven anti-resonant elastic modes identified above. In doing so, we choose frequencies which maximise the mechanical quality factor $Q_{\text{m}}$ of each mode, in line with the earlier discussion of the acoustic anti-resonance.

As shown in Table~\ref{tab:GainValues}, we predict  Brillouin gains on the order of $50~(\text{Wm})^{-1}$ dominated by the photo-elastic effect, with a greatest observed value for the 5th order elastic mode at a Brillouin shift of approximately 23~GHz. In achieving gains of this magnitude we have prioritised coupling with low-order optical modes. Several such modes are shown in Appendix~\ref{OpticalModes}. Higher-order optical modes typically produce greater Brillouin gains but have been excluded here in the interest of experimental practicality.

Note that only odd-numbered values of $l$ have been included in our calculations of BSBS gains. This is due to the particular symmetries associated with the transverse displacement profile $\modeprofile$ of even-numbered anti-resonant modes, resulting in vanishing photo-elastic and moving boundary contributions to gain, regardless of the coupled optical mode.


\begin{table}
    \begin{tabular}{ c c c | c c c }
        \hline\hline
        \multicolumn{3}{c}{\textbf{Elastics}} & \multicolumn{2}{c}{\textbf{Optics}} &                                                            \\
        \hline \hline
        $l$                                   & $\Omega/2\pi$ (GHz)                 & $Q_m$ & $\lambda$ (\micron) & $\Gamma$ (Wm)$^{-1}$ &       \\
        \hline \hline
                                              &                                     &       & 1.3                 & 28.5                 & (i)   \\
        5                                     & 23                                  & 490   & 1.2                 & 27.0                 & (ii)  \\
                                              &                                     &       & 1.1                 & 60.4                 & (iii) \\
        \hline
                                              &                                     &       & 1.1                 & 1.30                 & (iv)  \\
        7                                     & 30                                  & 560   & 1.0                 & 4.92                 & (v)   \\
                                              &                                     &       & 1.0                 & 37.0                 & (vi)  \\
        \hline
                                              &                                     &       & 1.1                 & 4.29                 &       \\
        9                                     & 36                                  & 740   & 1.0                 & 8.26                 &       \\
                                              &                                     &       & 0.99                & 15.6                 &       \\
    \end{tabular}
    \caption{Backward Stimulated Brillouin Scattering (BSBS) gain coefficients for the anti-resonant elastic modes and conventionally guided optical modes of a Si/SiO$_{2}$ rib waveguide. For each anti-resonant mode we have chosen the frequency $\Omega$ which maximises the mechanical quality factor $Q_{\text{m}}$. Optical wavelengths are provided for reference, and select mode profiles are available in Appendix~\ref{OpticalModes}.}
    \label{tab:GainValues}
\end{table}

An alternative formulation of this result, showing a series of BSBS gains for a fixed optical mode, over a range of detuning between the pump and Stokes frequencies, is shown in Appendix~\ref{app:LogGainPlot2}.


\section{Summary}
In this work we have built on the initial proposal of Anti-Resonant Reflecting Acoustic Waveguides (ARRAWs) \cite{schmidt_arraw_2020} for the co-localization of optical and elastic fields, moving from idealised one-dimensional cylindrical and planar geometries to a more experimentally-realizable family of rib-like silicon-on-insulator platforms. We predict that these structures are capable of supporting Brillouin gains on the order of $50~(\mathrm{Wm})^{-1}$ for BSBS --- a value comparable to the experimental demonstrations of BSBS in several other competing platforms.

The generality of the ARRAW concept suggests that anti-resonant guidance could be employed for high-$Q_{\mathrm{m}}$ elastic guidance in a multitude of material platforms, so long as the basic requirement of a multi-layered design configuration is met. Hence, future possibilities include the modeling of alternative waveguide designs which act to further suppress the dissipation of elastic waves, such as in suspended waveguides, and of unique material combinations, such as those with different velocity orderings to those present here.

\begin{acknowledgments}
    M.J.S acknowledges funding from the Australian Research Council Discovery Project scheme under projects DP200101893 and DP220100488. M.K.S. acknowledges funding from the Macquarie University Research Fellowship Scheme (MQRF0001036), and the Australian Research Council Discovery Early Career Researcher Award DE220101272. The authors acknowledge fruitful discussions with Christopher G. Poulton.
\end{acknowledgments}

\section*{Data Availability Statement}

COMSOL .mph file and Python scripts are available upon reasonable request.

\appendix

\section{Appendixes}

\subsection{Material properties}\label{MaterialConstants}
We model the structural response of both silicon (Si, with refractive index set to $n=3.48$) and silica (SiO$_2$, $n=1.45$) as cubic materials. The relevant material properties are shown below in Table.~(\ref{fig:MaterialValuesTable}). The photoelastic and stiffness tensor are taken from published work~\cite{weber_handbook_2018, dolbow_effect_1996}.


\begin{table}[ht]
    \footnotesize
    \begin{tabular}{ c c c c}
                          & \textbf{Density}      & \textbf{Stiffness tensor}                              & \textbf{Photo-elastic tensor}                    \\
        \textbf{Material} & \textbf{(kg/m$^{3}$)} & \textbf{($\boldsymbol{c_{11},~c_{12},~c_{44}}$)~(GPa)} & \textbf{($\boldsymbol{p_{11},~p_{12},~p_{44}}$)}
        \\
        \hline
        Silicon           & 2329                  & ($165.6,~63.9,~79.5)$                                  & ($-0.094,~0.017,~-0.051$)                        \\
        Silica            & 2203                  & ($78.6,~16.1,~31.2$)                                   & ($0.12,~0.27,~-0.075$)                           \\
        \hline \hline
    \end{tabular}
    \caption{The relevant material properties used to describe silicon (Si) and silica (SiO$_2$).}
    \label{fig:MaterialValuesTable}
\end{table}

\subsection{Meshing Details}\label{MeshingDetails}
We found that COMSOL was susceptible to identifying spurious surface-bound states at the substrate-PML boundary. While finer and more consistent meshing helped to alleviate these issues, doing so greatly impacts the computational cost of each simulation. Our compromise was to greatly increase mesh density in the core and cladding of the waveguide, where the fine structure of the field is most significant for our purposes, while also employing filtering conditions designed to ignore these artificially localised modes.

Another challenge was in the meshing of the Perfectly Matched Layer (PML). In particular, we found that the performance of the PML was highly sensitive to the structure of the mesh. Due to COMSOL's implementation method for artificial domains, periodic conditions require that the mesh is identical on the source and destination boundaries~\cite{zienkiewicz_novel_1983}. This will not ordinarily be true for \textsl{free} meshing methods, which don't tend to conserve the number of elements composing the boundaries of any given domain. The solution employed here was to use a \textsl{mapped distribution}, with a prescribed number of elements on each boundary. Further, we isolated the core and cladding of the model with a small \textsl{free triangular} mesh. This allows us to create vastly different element densities in the rib and substrate, providing more freedom to optimise the computational costs associated with solving the model.

\subsection{Numerical solver}\label{NumericalSolverDetails}
All calculations were conducted using the \textsl{Eigenfrequency study} in COMSOL's  \textsl{Structural Mechanics} module. This involves finding the frequencies $\Omega$ and transverse displacement profiles $\tilde{\vecu}(x, y)$ for a specified number of modes at a particular wavenumber $q$, about some centre frequency $\Omega_{0}$. By defining this centre frequency as a function of the free parameter $q$ such that:
\begin{align}
    \Omega_{0}(q) = \frac{qV_\text{S}^{\text{co}}}{n_{\text{eff}}}\,,
\end{align}
and specifying the value of $n_{\text{eff}}$, we can control whether the obtained solutions lie in the conventionally-guided or leaky regions of the dispersion diagram.

\subsection{Identifying conventionally guided modes}\label{app:tirmodes}
Generally, one would expect that the conventionally guided modes should appear in the spectrum unambiguously. While filtering is not typically required for conventionally guided modes, in the present work, we found that the PML did not perfectly enforce the open boundary condition, resulting in a continuum of non-physical modes which occupied the spatial same region of the dispersion plane, but which were extended across the whole simulation domain. These states needed to be removed to identify the well-defined curves of the TIR modes shown in the region $1\lesssim \neff\lesssim n_R$ seen in Fig.~\ref{fig:Dispersion}. This accounts for the occasional gaps in the red dispersion bands in that figure.
To do so, we filtered the raw set of eigenstates from the numerical solver based on the spatial distribution of the mode profile, $\modeprofile$ applying two conditions. Firstly, we require that the first-moment of the $y$ coordinate:
\begin{align}
    \langle y \rangle = \frac{\int y|\modeprofile|^{2} \, \dx\dy}{\int |\modeprofile|^{2}\, \dx\dy}\,,
\end{align}
describe modes localized to the cladding. Secondly we favored modes with  small \textsl{second moment widths}:
\begin{align}\label{SecondMomentWidth}
    w_{y}=\sqrt{\langle y^{2} \rangle - \langle y \rangle^{2}} < d\,.
\end{align}

\subsection{Optical modes}\label{OpticalModes}
\begin{figure}[ht]
    \begin{center}
        \includegraphics[width=0.7\linewidth]{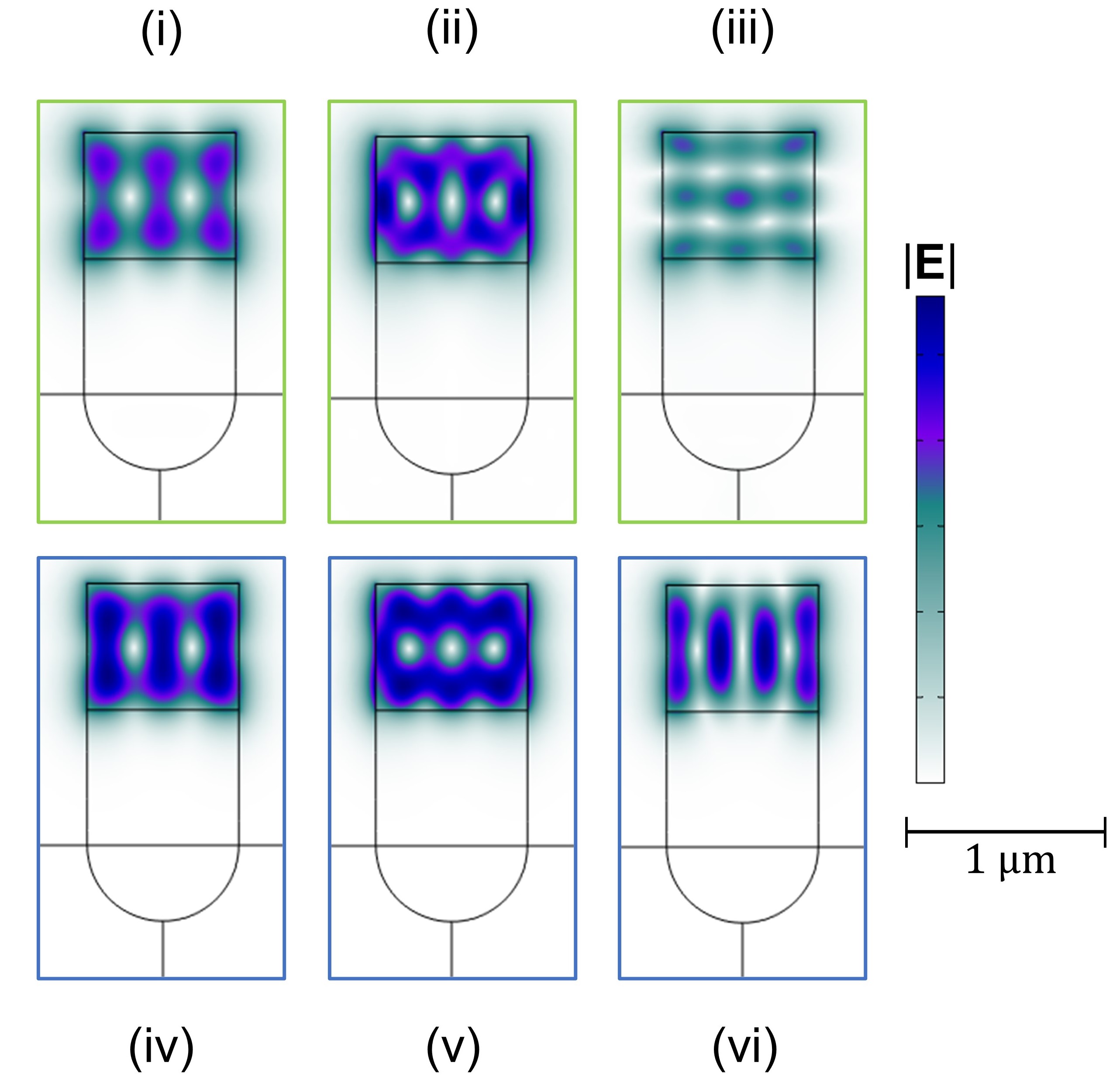}
        \caption{
            The mode profiles $\tilde{\boldsymbol{e}}(x, y)$ of those optical modes used to produce gains (i)-(vi) in Table~\ref{tab:GainValues}. Here, top and bottom optical modes are coupled with the $l=5$ or 7 anti-resonant mode, respectively.
        }
        \label{fig:OpticalModes}
    \end{center}
\end{figure}

Fig. ~\ref{fig:OpticalModes} shows the optical modes indicated in Table~\ref{tab:GainValues}, coupled to the 3rd (green) and 5th (blue) order anti-resonant modes. As mentioned in the main body of this text, we prioritised coupling to low-order optical modes, in the interest of experimental viability.

\subsection{Backward SBS calculations}\label{BSBSCalcs}
Here we summarise the  theoretical procedure used to calculate the SBS gain~\cite{wolff_stimulated_2015,sipe_hamiltonian_2016}.  Expanding the optical and elastic fields in waveguide modes, we have
\begin{align}\label{FormOfElectricField}
    \textbf{E} & = \sum_{n} a_{n}(z, t)\boldsymbol{e}_{n}(\boldsymbol{r}, t) + \mathrm{c.c.} \nonumber           \\
               & = \sum_{n} a_{n}(z, t)\tilde{\boldsymbol{e}}_{n}(x, y)e^{i(k_{n}z-\omega_{n}t)} + \mathrm{c.c.} \\
    \textbf{U} & = b(z, t)\boldsymbol{u}(\boldsymbol{r}, t) + \mathrm{c.c.} \nonumber                            \\
               & = b(z, t)\tilde{\boldsymbol{u}}(x, y)e^{i(qz-\Omega t)} + \mathrm{c.c.}\,,
\end{align}
where $a_{n}(z, t)$ is the complex-valued, dimensionless optical envelope function, $\tilde{\boldsymbol{e}}_{n}(x, y)$ is the mode profile containing the transverse dependence of the field, $k_{n}$ is the optical wavenumber and $\omega_{n}$ is a complex-valued angular frequency. Analogous quantities $b(z,t)$, $\boldsymbol{u}(\boldsymbol{r}, t)$ and $q$ are defined for the elastic field respectively. The summation in Eq.~(\ref{FormOfElectricField}) includes contributions of both the \textsl{pump} and $\textsl{Stokes}$ fields. For backwards SBS,the fields  occupy the same waveguide mode and we drop the mode index subscript. The field normalizations are expressed in terms of the modal electromagnetic power flux
\begin{align}
    \mathcal{P} & = 2\mathrm{Re}\int \hat{\boldsymbol{z}} \cdot (\tilde{\textbf{e}}\times\tilde{\textbf{h}}^{*})\,\mathrm{d}A\,,
\end{align}
and elastic energy density
\begin{align}
    \mathcal{E}_{b} & = \int \left[\rho|\partial_{t}\tilde{\textbf{u}}|^{2} + S_{ij}c_{ijkl}S_{kl}\right]\,\mathrm{d}A = 2\Omega^{2} \int \rho|\tilde{\textbf{u}}|^2\,\mathrm{d}A\,,
\end{align}
where the integration is over the whole transverse plane for the electromagnetic field, and over the waveguide profile for the elastic field. The photo-elastic contribution to the Stimulated Brillouin Scattering gain coefficient is
\begin{align}
    \mathcal{Q}^{\mathrm{PE}} = \varepsilon_{0}\varepsilon_{\mathrm{r}}^{2}\int \sum_{ijkl}[\tilde{e}_{i}(\boldsymbol{r})]^{*}{\tilde{e}}_{j}(\boldsymbol{r})p_{ijkl}(\boldsymbol{r})\boldsymbol{S}_{kl}^{*}(\boldsymbol{r})\,\mathrm{d}A\,.
\end{align}
Similarly for the moving boundary effect, we define
\begin{align}
    \mathcal{Q}^{\mathrm{MB}} = \int_{C} &
    (\hat{\textbf{n}}\cdot\tilde{\textbf{u}}^{*})
    \nonumber                                                                                                                                                                                                               \\
                                         & \times \Big[(\varepsilon_{a}-\varepsilon_{b})\varepsilon_{0}(\hat{\boldsymbol{n}}\times\tilde{\textbf{e}})^{*}\cdot(\hat{\boldsymbol{n}}\times\tilde{\textbf{e}}) \Big.\nonumber \\
                                         & \quad
        \Big. - (\varepsilon_{a}^{-1}-\varepsilon_{b}^{-1})\varepsilon_{0}^{-1}(\hat{\boldsymbol{n}}\cdot\tilde{\textbf{d}})^{*}(\hat{\boldsymbol{n}}\cdot\tilde{\textbf{d}})\Big]\,\mathrm{d}l\,,
\end{align}
where the integral runs over the contour $C$ that separates materials $a$ and $b$, corresponding to the core and its surrounding material(s), respectively. The  normal vector $\hat{\boldsymbol{n}}$ has been defined as outward-pointing. These two values are summed to give a total coupling coefficient $\mathcal{Q}_\text{SBS}$ given by Eq.~(\ref{CouplingCoef}), and yield the Backwards Stimulated Brillouin Scattering gain coefficient
\begin{align}
    \Gamma = 4\omega_{\text{P}}\frac{Q_{\text{m}}}{\mathcal{P}_{\text{P}}\mathcal{P}_{\text{S}}\mathcal{E}_{\text{m}}}|\mathcal{Q}_{\text{SBS}}|^{2}\,,
\end{align}
measured in units of (Wm)$^{-1}$.

\subsection{Gains for a fixed optical mode}\label{app:LogGainPlot2}
\begin{figure}[ht]
    \begin{center}
        \includegraphics[width=\linewidth]{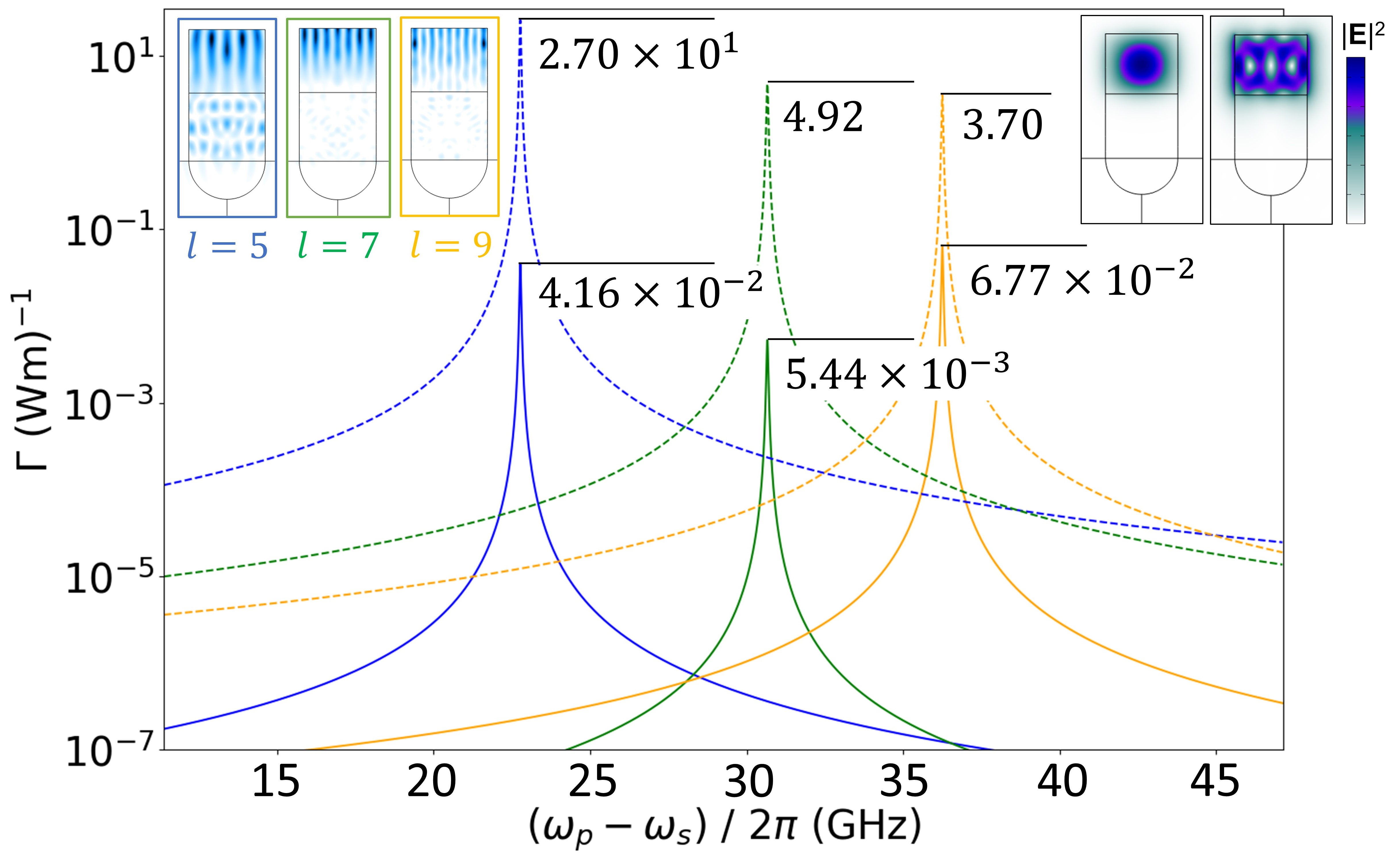}
        \caption{
            The BSBS gains in units (Wm)$^{-1}$ of a fixed optical mode for a range of detuning between the pump and Stokes frequencies. The solid lines correspond to those gains achieved using the fundamental mode.
        }
        \label{fig:LogGainPlot}
    \end{center}
\end{figure}

The BSBS gains for a fixed optical mode coupled to the 3rd, 4th and 5th order anti-resonant modes. This formulation of our results in Fig.~\ref{fig:LogGainPlot}, framed in reference to the optical rather than elastic quantities, is more commonly observed in the literature and so is included here for completeness. Here, we have searched for optical modes that satisfy the well-known phase-matching condition of BSBS, $q\approx k_{\mathrm{p}} - k_{\mathrm{S}} = 2k_{\mathrm{P}}$, for wavenumbers $k_{\mathrm{p}}$ and $k_{\mathrm{S}}$ of the pump and Stokes modes, respectively. From this condition, it must therefore also be the case that $\omega_{\mathrm{p}}-\omega_{\mathrm{S}} = 2\omega_{\mathrm{p}}$ is equal to the angular frequency of the elastic mode, $\Omega$.

\bibliography{articles}

\end{document}